\documentclass[prb,preprint,superscriptaddress]{revtex4}

\usepackage{graphicx}

\newcommand{\up}{\uparrow}
\newcommand{\dn}{\downarrow}

\begin{document}
\title{Spin Orbit Coupling and Spin Waves in Ultrathin Ferromagnets:
The Spin Wave Rashba Effect}
\author{A. T. Costa}
\affiliation{Instituto de F\'isica, Universidade Federal Fluminense, 
24210-340 Niter\'oi, RJ, Brasil.}
\author{R. B. Muniz}
\affiliation{Instituto de F\'isica, Universidade Federal Fluminense, 
24210-340 Niter\'oi, RJ, Brasil.}
\author{S. Lounis}
\affiliation{Department of Physics and Astronomy,
University of California Irvine, California, 92697, U. S. A.}
\author{A. B. Klautau}
\affiliation{Departamento de Fisica, Universidade Federal do Par\'a,
Bel\'em, PA, Brazil.}
\author{D. L. Mills}
\affiliation{Department of Physics and Astronomy,
University of California Irvine, California, 92697, U. S. A.}

\begin{abstract}
We present theoretical studies of the influence of spin orbit coupling on the
spin wave excitations of the Fe monolayer and bilayer on the W(110) surface. The
Dzyaloshinskii-Moriya interaction is active in such films, by virtue of the
absence of reflection symmetry in the plane of the film. When the magnetization
is in plane, this leads to a linear term in the spin wave dispersion relation
for propagation across the magnetization. The dispersion relation thus assumes a
form similar to that of an energy band of an electron trapped on a semiconductor
surfaces with Rashba coupling active. We also show SPEELS response functions
that illustrate the role of spin orbit coupling in such measurements. In
addition to the modifications of the dispersion relations for spin waves, the
presence of spin orbit coupling in the W substrate leads to a substantial
increase in the linewidth of the spin wave modes. The formalism we have
developed applies to a wide range of systems, and the particular system explored
in the numerical calculations provides us with an illustration of phenomena
which will be present in other ultrathin ferromagnet/substrate combinations.
\end{abstract}

\maketitle

\section{Introduction}
The study of spin dynamics in ultrathin ferromagnets is of fundamental interest,
since new physics arises in these materials that has no counterpart in bulk
magnetism. Examples are provided by relaxation mechanisms evident in
ferromagnetic resonance and Brillouin light scattering studies,\cite{1,2,3}
and also for the large wave vectors probed by spin polarized electron loss 
spectroscopy (SPEELS).\cite{4} Of course, by now the remarkable impact of 
ultrathin film structures on magnetic data storage is very well known, and other
applications that exploit spin dynamics in such materials are envisioned. Thus
these issues are important from a practical point of view as well as from that
of fundamental physics. 

Theoretical studies of the nature of spin waves in ultrathin films adsorbed on
metal substrates have been carried out for some years now, along with comparison
with descriptions provided with the Heisenberg model.\cite{5}  In this paper, we
extend the earlier theoretical treatments to include the influence of spin orbit
coupling on the spin wave spectrum of ultrathin films.
This extension is motivated by a most interesting discussion of the ground state
of the Mn monolayer on the W(110) surface. A nonrelativistic theoretical study
of this system predicted that the ground state would be antiferromagnetic in
character.\cite{7} This prediction was confirmed by spin polarized scanning
tunneling microscope studies of the system.\cite{8} However, recent experimental STM
data with a more sensitive instrument showed a more complex ground state,
wherein the ground state is in fact a spin density wave.\cite{9} One can construct
the new state by beginning with the antiferromagnet, and then superimposing on
this a long wavelength modulation on the direction of the moments on the
lattice. The authors of ref.~\onlinecite{9} argued that the lack of reflection symmetry of
the system in the plane of the film activates the Dzyaloshinskii Moriya (DM)
interaction, and the new state has its origin in this interaction.  They also
presented relativistic and ab initio calculations that gave an excellent account
of the new data. The reflection symmetry is broken simply by the presence of the
substrate upon which the film is grown. This argument to us is most intriguing,
since one can then conclude that the DM interaction must be active in any
ultrathin ferromagnet; the substrate is surely always present.  The DM
interaction has its origin in the spin orbit interaction, which of course is
generally very weak in magnets that incorporate the 3d transition elements as
the moment bearing entities. However, in the case of the Mn monolayer on W(110)
hybridization between the Mn 3d and the W 5d orbitals activates the very large W
spin orbit coupling, with the consequence that the strength of the DM
interaction can be substantial, as illustrated by the calculations presented in
ref.~\onlinecite{9}. One may expect to see substantial impact of the DM interaction in
other ultrathin magnets grown on 5d substrates, and possibly 4d substrates as
well.

We have here another example of new physics present in ultrathin magnets that is
not encountered in the bulk form of the material from which the ultrathin
structure is fabricated. The purpose of this paper is to present our theoretical
studies of spin orbit effects on spin waves and also on the dynamic
susceptibility of a much studied ultrathin film/substrate combination, the Fe
monolayer and bilayer on W(110). We find striking effects. For instance, when
the magnetization is in plane, as we shall see the DM interaction introduces a
term linear in wave vector in the dispersion relation of spin waves. Thus the
uniform spin wave mode at zero wave vector acquires a finite group velocity. We
find this to be in the range of $2\times 10^5$ cm/sec for the Fe monolayer on W(110).
Furthermore, left/right asymmetries appear in the SPEELS response functions.
Thus, we shall see that spin orbit coupling has clear effects on the spin
excitations of transition metal ultrathin ferromagnets grown on 5d substrates.

We comment briefly on the philosophy of the approach used here, and in various
earlier publications.\cite{5} Numerous authors proceed as follows. One may generate
a description of the magnetic ground state of the adsorbed films by means of an
electronic structure calculation based on density functional theory. It is then
possible to calculate, within the framework of an adiabatic approximation,
effective Heisenberg exchange integrals Jij  between the magnetic moments in
unit cell i and unit cell j. These may be entered into a Heisenberg Hamiltonian,
and then spin wave dispersion relations may be calculated through use of spin
wave theory. It has been known for decades\cite{10} that in the itinerant 3d
magnets, effective exchange interactions calculated in such a manner have very
long range in real space. Thus, one must include a very large number of distant
neighbors in order to obtain converged results. This is very demanding to do
with high accuracy for the very numerous distant neighbors, since the exchange
interactions become very small as one moves out into distant neighbor shells.

At a more fundamental level, as noted briefly above, discussions in earlier
publications show that in systems such as we study here, the adiabatic
approximation breaks down badly, with qualitative consequences.\cite{5} First, spin
wave modes of finite wave vector have very short lifetimes, by virtue of decay
into the continuum of particle hole pairs (Stoner excitations) even at the
absolute zero of temperature\cite{5,11} whereas in Heisenberg model descriptions
their lifetime is infinite. In multi layer films, the earlier calculations show
that as a consequence of the short lifetime, the spectrum of spin fluctuations
at large wave vectors contains a single broad feature which disperses with wave
vector in a manner similar to that of a spin wave; this is consistent with
SPEELS data on an eight layer film of Co on Cu(100).\cite{4} This picture stands in
contrast to that offered by the Heisenberg model, in which a film of N layers
has N spin wave modes for each wave vector, and each mode has infinite lifetime. 

The method developed earlier, and extended here to incorporate spin orbit
coupling, takes due account of the breakdown of the adiabatic approximation and
also circumvents the need to calculate effective exchange interactions in real
space out to distant neighbor shells. We work directly in wave vector space
through study of the wave vector and frequency dependent susceptibility
discussed below, denoted as
$\chi_{+,-}(\vec{Q}_\parallel,\Omega;l_\perp,l'_\perp)$. The imaginary part of
this object, evaluated for $l_\perp=l'_\perp$ and considered as a function of
frequency $\Omega$ for fixed wave vector $\vec{Q}_\parallel$ provides us with
the frequency spectrum of spin fluctuations on layer $l_\perp$ for the wave
vector chosen. Spin waves appear as peaks in this function, very much as they do
in SPEELS data, and in a manner very similar to that used by experimentalist we
extract a dispersion relation for spin waves by following the wave vector
dependence of the peak frequency. We never need to resort to a real space
summation procedure over large number of neighbors, coupled by very tiny
exchange couplings. The spin wave exchange stiffness can be extracted either by
fitting the small wave vector limit of the dispersion relation so determined, or
alternatively by utilizing an expression derived earlier\cite{5} which once again
does not require a summation in real space.  

We comment on another feature of the present study. In earlier 
calculations,\cite{5,11,14} as in the present paper, an empirical tight binding description
forms the basis for our description of the electronic structure. Within this
approach, referred to as a multi band Hubbard model, we can generate the wave
vector and frequency dependent susceptibility for large systems. In the earlier papers,
effective tight binding parameters were extracted from bulk electronic structure
calculations. The present studies are based on tight binding parameters obtained
directly from a RS-LMTO-ASA calculation for the Fe/W(110) system. We also obtain
tight binding parameters by fitting KKR based electronic structure calculations
for the ultrathin film/substrate combinations of interest. We find that spin
waves in the Fe/W(110) system are quite sensitive to the empirical tight binding
parameters which are employed, though as we shall see the various descriptions
provide very similar pictures of the one electron local density of states. 

We note that Udvardi and Szunyogh\cite{12} have also discussed the influence of spin
orbit coupling on the dispersion relation of spin waves in the Fe monolayer on
W(110) within the framework of the adiabatic approach discussed above, where
exchange interactions and other magnetic parameters are calculated in real
space. We shall discuss a comparison with our results and theirs below. There
are differences. Most particularly, we note that in Fig. 3, the authors of 
ref.~\onlinecite{12} provide two dispersion curves for propagation perpendicular to the
magnetization, whereas in a film such as this with one spin per unit cell there
can be only one magnon branch. Additionally and very recently, Bergmann and
coworkers\cite{45} investigated within an adiabatic approach finite temperature
effects on the magnon spectrum of Fe/W(110). 

In section II, we comment on our means of introducing spin orbit coupling into
the theory. The results of our calculations are summarized in Section III and
concluding remarks are found in section IV.

\section{Calculation of the Dynamic Susceptibility in the Presence of Spin Orbit Coupling}

The formalism for including spin orbit coupling effects in our description of
spin dynamics is quite involved, so in this section we confine our attention to
an outline of the key steps, and an exposition of the overall structure of the
theory. Our starting point is the multi band Hubbard model of the system that
was employed in our earlier study of spin dynamics in ultrathin ferromagnets.
The starting Hamiltonian is written as\cite{5}
\begin{equation}
H=\sum_{ij}\sum_{\mu\nu\sigma}T_{ij}^{\mu\nu}c_{i\mu\sigma}^{\dagger}c_{j\nu\sigma}+
\frac{1}{2}\sum_{\mu\nu\mu'\nu'}\sum_{i\sigma\sigma'}
U_{i;\mu\nu,\mu'\nu'}
c_{i\mu\sigma}^{\dagger}c_{i\nu\sigma'}^{\dagger}c_{i\nu'\sigma'}c_{i\mu'\sigma}
\label{hamilt0}
\end{equation}
where $i$ and $j$ are site indices, $\sigma,\sigma'$ refer to spin, 
and $\mu,\nu$ to the tight binding orbitals, nine in number for each site, 
which are included in our treatment. The Coulomb interactions operate only
within the 3d orbitals on a given lattice site. The film, within which
ferromagnetism is driven by the Coulomb interactions, sits on a semi-infinite
substrate within which the Coulomb interaction is ignored. 

In our empirical tight binding picture, the spin orbit interaction adds a term
we write as
\begin{equation}
H_{SO}=\sum_{i}\sum_{\mu\nu}\frac{\lambda_{i}}{2}
\left[ L_{\mu\nu}^{z}(c_{i\mu\uparrow}^{\dagger}c_{i\nu\uparrow}-
c_{i\mu}^{\dagger}c_{i\nu\downarrow})+
L_{\mu\nu}^{+}c_{i\mu\downarrow}^{\dagger}c_{i\nu\uparrow}+
L_{\mu\nu}^{-}c_{i\mu\uparrow}^{\dagger}c_{i\nu\downarrow}\right]
\label{hamilt_SO}
\end{equation}
where $\vec{L}$ is the angular momentum operator, $\lambda_i$ is the local
spin-orbit coupling constant, $L^\pm=L^x\pm iL^y$ and
$L^\alpha_{\mu\nu}=\langle\mu|L^\alpha|\nu\rangle$. 
We assume that the spin orbit interaction, present both within the ferromagnetic
film and the substrate, operates only within the 3d atomic orbitals.
A convenient tabulation of matrix elements of the
orbital angular momentum operators is found in ref.~\onlinecite{13}.

Information on the spin waves follows from the study of the spectral density of
the transverse dynamic susceptibility 
$\chi_{+,-}(\vec{Q}_\parallel,\Omega;l_\perp,l'_\perp)$ as discussed above. From the text around
Eq. (1) of ref.~\onlinecite{14}, we see that this function describes the amplitude of the
transverse spin motion (the expectation value of the spin operator $S^+$ in the
layer labeled $l_\perp$) to a fictitious transverse magnetic field of frequency
$\Omega$ and wave vector $\vec{Q}_\parallel$ parallel to the film surface
that is applied to layer $l'_\perp$ of the
sample. The spectral density, given by  
$\mathrm{Im}\{\chi_{+,-}(\vec{Q}_\parallel,\Omega;l_\perp,l'_\perp)\}$, 
when multiplied by the Bose Einstein
function $n(\Omega)=[\exp(\beta\Omega)-1]^{-1}$
 is also the amplitude of thermal spin fluctuations of wave vector $\vec{Q}_\parallel$
and frequency $\Omega$ in layer $l_\perp$. We obtain information regarding the character
(frequency, linewidth, and amplitude in layer $l_\perp$) of spin waves from the study of this
function, as discussed earlier.\cite{5} 

Our previous analyses are based on the study of the dynamic susceptibility just
described through use of the random phase approximation (RPA) of many body
theory. The Feynman diagrams included in this method are the same as those
incorporated into time dependent density functional theory, though use of our
Hubbard model allows us to solve the resulting equation easily once the very
large array of irreducible particle hole propagators are generated numerically. 

Our task in the present paper is to extend the RPA treatment to incorporate spin
orbit coupling. The extension is non trivial. The quantity of interest, referred
to in abbreviated notation as  $\chi_{+,-}$, may be expressed as a commutator of the spin
operators $S^+$ and $S^-$ whose precise definition is given earlier.\cite{5,12} 
With spin orbit coupling ignored, the RPA decoupling procedure leads to a closed equation
for $\chi_{+,-}$. When the RPA decoupling is carried out in its presence, we are led to a
sequence of four coupled equations which include new objects we may refer to as $\chi_{-,-}$,
$\chi_{\uparrow,-}$ and $\chi_{\downarrow,-}$. The number of irreducible particle hole propagators that must be
computed likewise is increased by a factor of four. For a very simple version of
a one band Hubbard model, and for a very different purpose, Fulde and Luther
carried out an equivalent procedure many years ago\cite{15}. In what follows, we
provide a summary of key steps along with expressions for the final set of
equations. 

To generate the equation of motion, we need the commutator of the operator
$S_{\mu\nu}^{+}(l,l')=c_{l\mu\uparrow}^{\dagger}c_{l'\nu\downarrow}$
with the Hamiltonian. One finds
\begin{equation}
[S_{\mu\nu}^{+}(l,l'),H_{SO}]=\frac{1}{2}\sum_{\eta}\{
\lambda_{l'}L^{+}_{\nu\eta}c^\dagger_{l\mu\uparrow}c_{l'\eta\uparrow}
-\lambda_l L^+_{\eta\mu}c^\dagger_{l\eta\downarrow}c_{l'\nu\downarrow}
+\lambda_{l'}L^z_{\nu\eta}c^\dagger_{l\mu\uparrow}c_{l'\eta\downarrow}
-\lambda_l L^z_{\eta\mu}c^\dagger_{l\eta\uparrow}c_{l'\nu\downarrow}\}.
\label{comm}
\end{equation}
The last two terms on the right hand side of Eq.~\ref{comm} lead to terms in the
equation of motion which involve $\chi_{+,-}$ whereas the first two terms couple
us to the entities $\chi_{\uparrow,-}$ and $\chi_{\downarrow,-}$. When we write
down the commutator of these new correlation functions with the spin orbit
Hamiltonian, we are led to terms which couple into the function $\chi_{-,-}$
which is formed from the commutator of two $S^-$ operators.  In the
absence of spin orbit coupling, a consequence of spin rotation invariance of the
Hamiltonian is that the three new functions just encountered vanish. But they do
not in its presence, and they must be incorporated into the analysis. 

One then introduces the influence of the Coulomb interaction into the equation
of motion, and carries out an RPA decoupling of the resulting terms. The
analysis is very lengthy, so here we summarize only the structure that results
from this procedure. Definitions of the various quantities that enter are given
in the Appendix. We express the equations of motion in terms of a $4\times 4$ 
matrix structure, where in schematic notation we let $\chi^{(1)}=\chi_{+,-}$, 
$\chi^{(2)}=\chi_{\uparrow,-}$, $\chi^{(3)}=\chi_{\downarrow,-}$ and 
$\chi^{(4)}=\chi_{-,-}$. The four coupled equations then have the form
\begin{equation}
	\Omega\chi^{(s)} = A^{(s)} + \sum_{s'}(B^{ss'}+\tilde{B}^{ss'})\chi^{(s')}
	\label{eom_RPA_1}
\end{equation}
Each quantity in Eq.~\ref{eom_RPA_1} has attached to it four orbital indices, 
and four site indices. To be explicit, $\chi^{(2)}=\chi_{\uparrow,-}$ 
which enters Eq. (4) is formed from the commutator of
the operator $c^\dagger_{l\mu\uparrow}c_{l'\nu\uparrow}$  with
$c^\dagger_{m\mu'\downarrow}c_{m'\nu'\uparrow}$ and in full we denote 
this quantity as $\chi^{(2)}_{\mu\nu;\mu'\nu'}(ll';mm')$. The site indices
label the planes in the film, and we suppress reference to $\Omega$  and
$\vec{Q}_\parallel$. The products
on the right hand side of Eq.~\ref{eom_RPA_1} are matrix multiplications that involve these
various indices. For instance, the object $\sum_{s'}B^{ss'}\chi^{(s')}$ is labeled 
by four orbital and four site indices so
\begin{equation}
	[B^{ss'}\chi^{(s')}]_{\mu\nu,\mu'\nu'}(ll';mm') = 
	\sum_{\gamma\delta}\sum_{nn'}B^{ss'}_{\mu\nu,\gamma\delta}(ll';nn')
	\chi^{(s')}_{\gamma\delta,\mu'\nu'}(nn';mm').
\end{equation}

One proceeds by writing Eq.~\ref{eom_RPA_1} in terms of the dynamic susceptibilities that
characterize the non-interacting system. These, referred to also as the
irreducible particle hole propagators, are generated by evaluating the
commutators which enter into the definition of $\chi^{(s)}$  in the non interacting ground
state.  These objects, denoted by $\chi^{(0s)}$ obey a structure similar to
Eq.~\ref{eom_RPA_1},
\begin{equation}
	\Omega\chi^{(0s)} = A^{(s)} + \sum_{s'}B^{ss'}\chi^{(s')}.
	\label{eom_HF}
\end{equation}
It is then possible to relate $\chi^{(s)}$ to $\chi^{(0s)}$ through 
the relation, using four vector notation,
\begin{equation}
	\vec{\chi}(\Omega) = \vec{\chi}^{(0)}(\Omega) + 
	(\Omega-B)^{-1}\tilde{B}\vec{\chi}(\Omega).
	\label{dyson}
\end{equation}
The matrix structure $\Gamma\equiv (\Omega-B)^{-1}$  may be generated from the
definition of $B$, which may be obtained from the equation of motion of the
non-interacting susceptibility, Eq.~\ref{eom_HF}. Then $\tilde{B}$ follows from the equation of
motion of the full susceptibility, as generated in the RPA. One may solve
Eq.~\ref{dyson}
\begin{equation}
	\vec{\chi}(\Omega) = 
	[I - (\Omega-B)^{-1}\bar{B}]^{-1}\vec{\chi}^{(0)}(\Omega),
\label{chi_RPA}
\end{equation}
so our basic task is to compute the non interacting susceptibility matrix
$\vec{\chi}^{(0)}$ and then carry out the matrix inversion operation displayed
in Eq.~\ref{chi_RPA}. For this we require the single particle Green’s functions (SPGFs)
associated with our approach.

To generate the SPGFs, we set up an effective single particle Hamiltonian
$H_{sp}$ by introducing a mean field approximation for the Coulomb interaction.
The general structure of the single particle Hamiltonian is
\begin{equation}
H_{sp}=\sum_{ij}\sum_{\mu\nu\sigma}
\tilde{T}_{ij}^{\mu\nu\sigma}c^\dagger_{i\mu\sigma}c_{j\nu\sigma} +
\sum_i\sum_{\mu\nu}\{
\alpha_{i;\mu\nu}^*c^\dagger_{i\mu\downarrow}c_{i\nu\uparrow}+
\alpha_{i;\mu\nu}c^\dagger_{i\mu\uparrow}c_{i\nu\downarrow}\}
\label{H_sp}
\end{equation}
where the effective hopping integral $\tilde{T}_{ij}^{\mu\nu\sigma}$ contains
the spin diagonal portion of the spin orbit interaction, along with the mean
field contributions from the Coulomb interaction. The form we use for the latter
is stated below. The coefficients in the spin flip terms are given by
\begin{equation}
\alpha_{i;\mu\nu} = \lambda_i L^-_{\mu\nu} - 
\sum_{\eta\gamma}U_{i;\eta\mu,\nu\gamma}\langle
c^\dagger_{i\eta\downarrow}c_{i\gamma\uparrow}\rangle .
\label{alphas}
\end{equation}
We then have the eigenvalue equation that generates the single particle
eigenvalues and eigenfunctions in the form
$H_{sp}|\phi_s\rangle=E_s|\phi_s\rangle$; we can write this in the explicit
form
\begin{equation}
\sum_l\sum_{\eta\sigma'}\left[ 
	 \delta_{\sigma\sigma'}\tilde{T}^{\mu\eta\sigma'}_{il} + 
\delta_{il}(\delta_{\sigma'\downarrow}\delta_{\sigma\uparrow}\alpha^*_{l;\mu\eta} +
\delta_{\sigma'\uparrow}\delta_{\sigma\downarrow}\alpha_{l;\mu\eta}) \right] 
\langle l\eta\sigma'|\phi_s\rangle = E_s\langle i\mu\sigma|\phi_s \rangle .
	\label{eigenvals}
\end{equation}
The single particle Green’s function may be expressed in terms of the quantities
that enter Eq.~\ref{eigenvals}. We have for this object the definition
\begin{equation}
	G_{i\mu\sigma;j\nu\sigma'}(t) = 
-i\theta(t)\langle \{ c_{i\mu\sigma}(t),c^\dagger_{j\nu\sigma'}(0)\}\rangle
\end{equation}
and one has the representation
\begin{equation}
	G_{i\mu\sigma;j\nu\sigma'}(\Omega) = 
	\sum_s \frac{\langle i\mu\sigma|\phi_s\rangle\langle\phi_s|j\nu\sigma'\rangle}
	{\Omega - E_s + i\eta} .
\end{equation}
These functions may be constructed directly from their equations of motion,
which read, after Fourier transforming with respect to time, 
\begin{equation}
	 -\sum_l\sum_{\eta\sigma''}\left[ 
	 \delta_{\sigma\sigma''}\tilde{T}^{\mu\eta\sigma''}_{il} + 
\delta_{il}(\delta_{\sigma''\downarrow}\delta_{\sigma\uparrow}\alpha^*_{l;\mu\eta} +
	    \delta_{\sigma''\uparrow}\delta_{\sigma\downarrow}\alpha_{l;\mu\eta})
	 \right] G_{l\eta\sigma'';j\nu\sigma'} + 
	\Omega G_{i\mu\sigma;j\nu\sigma'} 
= \delta_{\sigma\sigma'}\delta_{\mu\nu}\delta_{ij}.
\label{spgf_eom}
\end{equation}
For the case where the substrate is semi infinite, our means of generating a
numerical solution to the hierarchy of equations stated in Eq.~\ref{spgf_eom} 
has been discussed earlier. What remains is to describe how the Coulomb interaction
enters the effective hopping integrals $\tilde{T}^{\mu\nu\sigma}_{ij}$  
that appear in Eq.~\ref{H_sp}, Eq.~\ref{eigenvals} and Eq.~\ref{spgf_eom}.

There are, of course, a large number of Coulomb matrix elements in the original
Hamiltonian, even if the Coulomb interactions are confined to within the 3d
shell. Through the use of group theory,\cite{17} the complete set of Coulomb matrix
elements may be expressed in terms of three parameters. These are given in Table
I of the first cited paper in ref.~\onlinecite{5}. In subsequent work, 
we have found that a much simpler structure\cite{18} nicely reproduces 
results obtained with the full three parameter form. We use the simpler 
one parameter form here, for which
$U_{i;\mu\nu,\mu'\nu'}=U_i\delta_{\mu\nu'}\delta_{\mu'\nu}$. 
Then in the mean field approximation, the Coulomb contribution to the single
particle Hamiltonian assumes the form
\begin{equation}
	H_{sp}^{(C)} = 
	-\sum_i\frac{U_i m_i}{2}\sum_{\mu}( c^\dagger_{i\mu\uparrow}c_{i\mu\uparrow} 
- c^\dagger_{i\mu\downarrow}c_{i\mu\downarrow})
	\label{U_HF}
\end{equation}
Here $m_i$ is the magnitude of the moment on site $i$. The Coulomb interactions
$U_i$  are
non zero only within the ultrathin ferromagnet, and the moments $m_i$, determined
self consistently, vary from layer to layer when we consider multi layer
ferromagnetic films. 

It should be noted that when the Ansatz just described is employed in
Eq.~\ref{alphas},
the term from the Coulomb interaction on the right hand side becomes
proportional to the transverse component of the moment located on site $i$ and
this vanishes identically. Thus, despite the complexity introduced by the spin
orbit coupling, when the simple one parameter Ansatz for the Coulomb matrix
elements is employed, one needs no parameters beyond the moment on each layer in
the self consistent loops that describe the ground state. In the present
context, this is an extraordinarily large savings in computational labor, and
this will allow us to address very large systems in the future. It is the case
that certain off diagonal elements such as $\langle
c^\dagger_{m\mu'\downarrow}c_{l'\nu\uparrow}\rangle$ appear in the quantities defined in
the Appendix. Notice, for example, the expressions in Eqs.~\ref{As}. Once the ground
state single particle Green’s functions are determined, such expectation values
are readily computed.

\section{Results and Discussion}

In earlier studies of Fe layers on W(110),\cite{19,20} as noted above, the
electronic structure was generated through use of tight binding parameters
obtained from bulk electronic structure calculations. These calculations
generate effective exchange interactions comparable in magnitude to those found
in the bulk transition metals,\cite{20} with the consequence that for both monolayer
Fe and bilayer Fe on W(110) the large wave vector spin waves generated by theory
are very much stiffer than found experimentally\cite{21,22} though it should be
noted that for the bilayer, the calculated value of the spin wave exchange
stiffness is in excellent accord with the data.\cite{23} Subsequent calculations
which construct the spin wave dispersion relation from adiabatic theory based on
calculations of effective exchange integrals also generate spin waves for the
monolayer substantially stiffer than found experimentally,\cite{12} though they are
softer than in our earlier work by a factor of two or so. We remark that it has
been suggested that the remarkably soft spin waves found experimentally may have
origin in carbon contamination of the monolayer and bilayer.\cite{20} We remark here
that this can be introduced during the SPEELS measurement. We note that the
magnetic properties of Fe monolayers grown on carbon free W(110)\cite{24} differ
dramatically from those grown on surfaces now known to be contaminated by 
carbon.\cite{25} In the former case, the domain walls have a thickness of 2.15
nm,\cite{24}
whereas in the latter circumstance very narrow walls with thickness bounded from
above by 0.6 nm are found.\cite{25} This suggests that the strength of the effective
exchange is very different in the two cases, with stiffer exchange in the carbon
free samples.

The considerations of the previous paragraph have motivated us to carry out a
series of studies of the effective exchange in the Fe monolayer on W(110) within
the framework of three different electronic structure calculations. We find that
although all three give local density of states that are very similar, along
with very similar energy bands when these are examined, the intersite exchange
interactions vary substantially. First, we have employed the parameter set
used earlier that is based on bulk electronic structures\cite{19,20} in new
calculations we call case A. In case B, we have employed an approach very
similar to that used in ref.~\onlinecite{12}, though in what follows our calculation of
effective exchange integrals is non relativistic. This is the Korringa Kohn
Rostoker Green’s Function (KKR-GF) method,\cite{26} which employs the atomic sphere
approximation and makes use of the Dyson equation $G=g+gVG$ as given in matrix notation.
This allows us to calculate the Green’s function $G$ of an arbitrary complex
system given the perturbing potential $V$ and the Green’s function $g$ of a
reference unperturbed system.  Within the Local Spin Density Approximation
(LSDA),\cite{27} We consider a slab of five monolayers of W with the experimental
lattice constant on top of which an Fe monolayer is deposited and relaxed by
-12.9\%\cite{12} with respect to the W  interlayer distance. Angular momenta up to
$l_{\mathrm{max}}=3$ were included in the Green’s functions with a $k$ mesh of
6400 points in the full two dimensional Brillouin zone. The effective exchange
interactions were calculated within the approximation of infinitesmal
rotations\cite{28} that allows one to use the magnetic force theorem. This
states that the energy change due to infinitesmal rotations in the moment
directions can be calculated through the Kohn Sham eigenvalues. 

Method C is the Real Space Linear-Muffin-Tin-Orbital approach as implemented,
also, in the atomic sphere approximation (RS-LMTO-ASA).\cite{29,30,31,32,33}
Due to its linear scaling, this method allows one to address the
electronic structure of systems with a large number of atoms for which the basic
eigenvalue problem is solved in real space using the Haydock recursion method.
The Fe overlayer on the W(110) substrate was simulated by a large bcc slab which
contained $\sim$6800 atoms, arranged in 12 atomic planes parallel to the (110)
surface, with the experimental lattice parameter of bulk W. One empty sphere
overlayer is included, and self consistent potential parameters were obtained
for the empty sphere overlayer, the Fe monolayer, and the three W layers
underneath using LSDA.\cite{34} For deeper W layers we use bulk potential. Nine
orbitals per site (the five 3d and 4 sp complex) were used to describe the Fe
valence band and the empty sphere overlayer, and for W the fully occupied 4f
orbitals were also included in the core. To evaluate the orbital moments we use
a scalar relativistic (SR) approach and include a spin orbit coupling term
$\lambda\vec{L}\cdot\vec{S}$  at
each variational step.\cite{35} In the recursion method the continued fraction has
been terminated after 30 recursion levels with the Beer Pettifor
terminator.\cite{36} The TB parameters so obtained are inserted into our semi-empirical scheme
and this allows us to generate the non interacting susceptibilities which enter
our full RPA description of the response of the structure. 

\begin{figure}
\includegraphics[width=0.4\textwidth,clip]{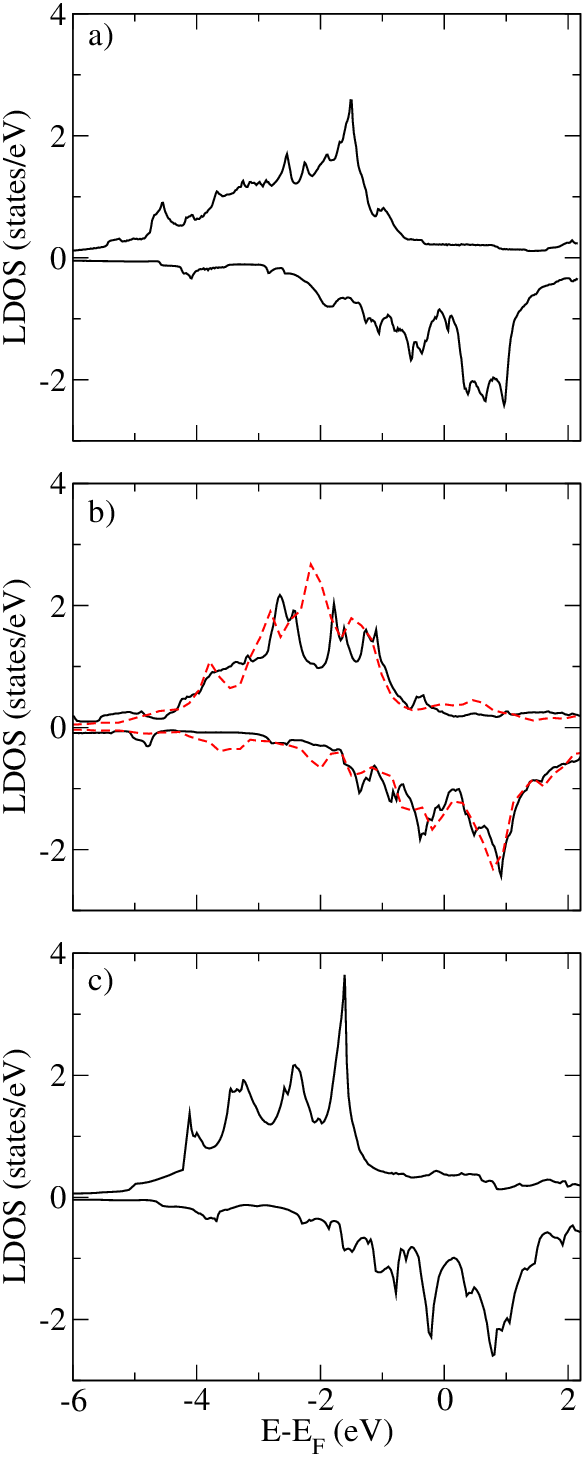}
\caption{(color online) For the Fe monolayer adsorbed on W(110), we show the local density of
states in the Fe monolayer. The majority spin density of states is shown
positive and the minority spin density of states is negative. The zero of energy
is at the Fermi energy. In (a), bulk electronic structure parameters are used as
in the second of the two papers cited in~\onlinecite{5} (caseA). In (b), we have the
density of states generated by method B. The black curve is found by fitting the
KKR electronic energy band structure to tight binding parameters as described in
the text, and the red curve is calculated directly from the KKR calculation. In
(c) we have the local density of states generated by method C.}
\label{fig1}
\end{figure}

In order to compare the electronic structures generated by the approaches just
described, we turn our attention to the local density of states for the majority
and minority spins in the adsorbed Fe monolayer. These are summarized in
Fig.~\ref{fig1}. 

The local densities of states (LDOS) generated by the three sets of TB
parameters have approximately the same overall features, as we see from
Fig.~\ref{fig1}.
The main differences appear in the majority spin band, which overlaps the 5d
states in the W substrate over a larger energy range than the minority band.
This is also true if we compare the LDOS generated by the tight binding
parameters extracted from the KKR electronic structure to the LDOS obtained
directly from the KKR calculations (red dashed line in Fig.~\ref{fig1}b). The Fe-W
hopping parameters are indeed the least accurate portion of our parametrization
scheme. In case A we just used the Fe-Fe bulk parameters to describe the Fe-W
hopping. In case B we extracted TB parameters for Fe by fitting a KKR
calculation of an unsupported Fe monolayer with a lattice parameter matching
that of the W substrate. For the Fe-W hopping we used the Fe parameters obtained
from the fitting, scaled to mimic the Fe-W distance relaxation.  The relaxation
parameter was chosen to give the correct spin magnetic moment for the adsorbed
Fe monolayer. In case C all parameters were directly provided by the RS-LMTO-ASA
code, but in the DFT calculations the Fe-W distance was assumed to be equal to
the distance between W layers. Thus, the main difference between cases B and C
is the treatment of the mixing between Fe and W states and this is expected to
affect more strongly the states that occupy the same energy range.

As noted above, while the local density of states provided by the three
approaches to the electronic structure are quite similar as we see from
Fig.~\ref{fig1} (and the same is true of the electronic energy bands themselves
if these are examined), the exchange interactions differ substantially for the
three cases. For the first, second and third neighbors we have (in meV) 42.5,
3.72 and 0.46 for model A, 28.7, -7.87 and 0.31, for model B and 11.23, -7.31 and
0.22 for model C. The authors of ref.~\onlinecite{12} find 10.84, -3.34 and 3.64 for these
exchange integrals. 

We now turn to our studies of spin excitations in the Fe monolayer and Fe
bilayer on W(110) within the framework of the electronic structure generated
through use of the approach in case C. We will discuss the influence of spin
orbit coupling on both the transverse wave vector dependent susceptibility
though study of the spectral density function
$A(\vec{Q}_\parallel,\Omega;l_\perp)=
-\mathrm{Im}\{\chi_{+,-}(\vec{Q}_\parallel,\Omega;l_\perp)\}$  
discussed in section I. This
function, for fixed wave vector $\vec{Q}_\parallel$, when considered as a 
function of frequency $\Omega$,
describes the frequency spectrum of the fluctuations of wave vector $\vec{Q}_\parallel$ of the
transverse magnetic moment in layer $l_\perp$  as noted above. In the frequency regime
where spin waves are encountered, this function is closely related to (but not
identical to) the response function probed in a SPEELS measurement. 

\begin{figure}
\includegraphics[width=0.45\textwidth,clip]{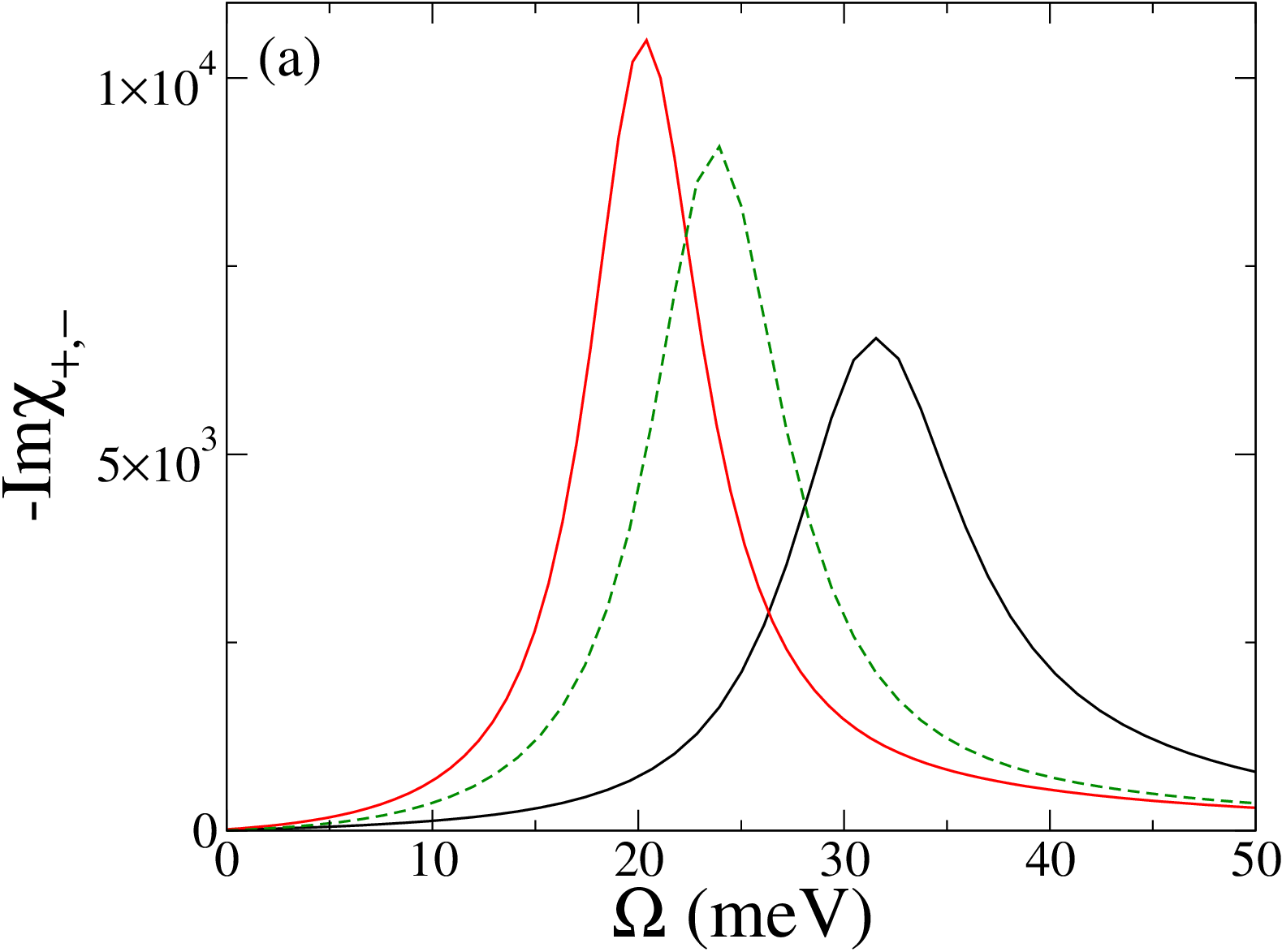}
\includegraphics[width=0.45\textwidth,clip]{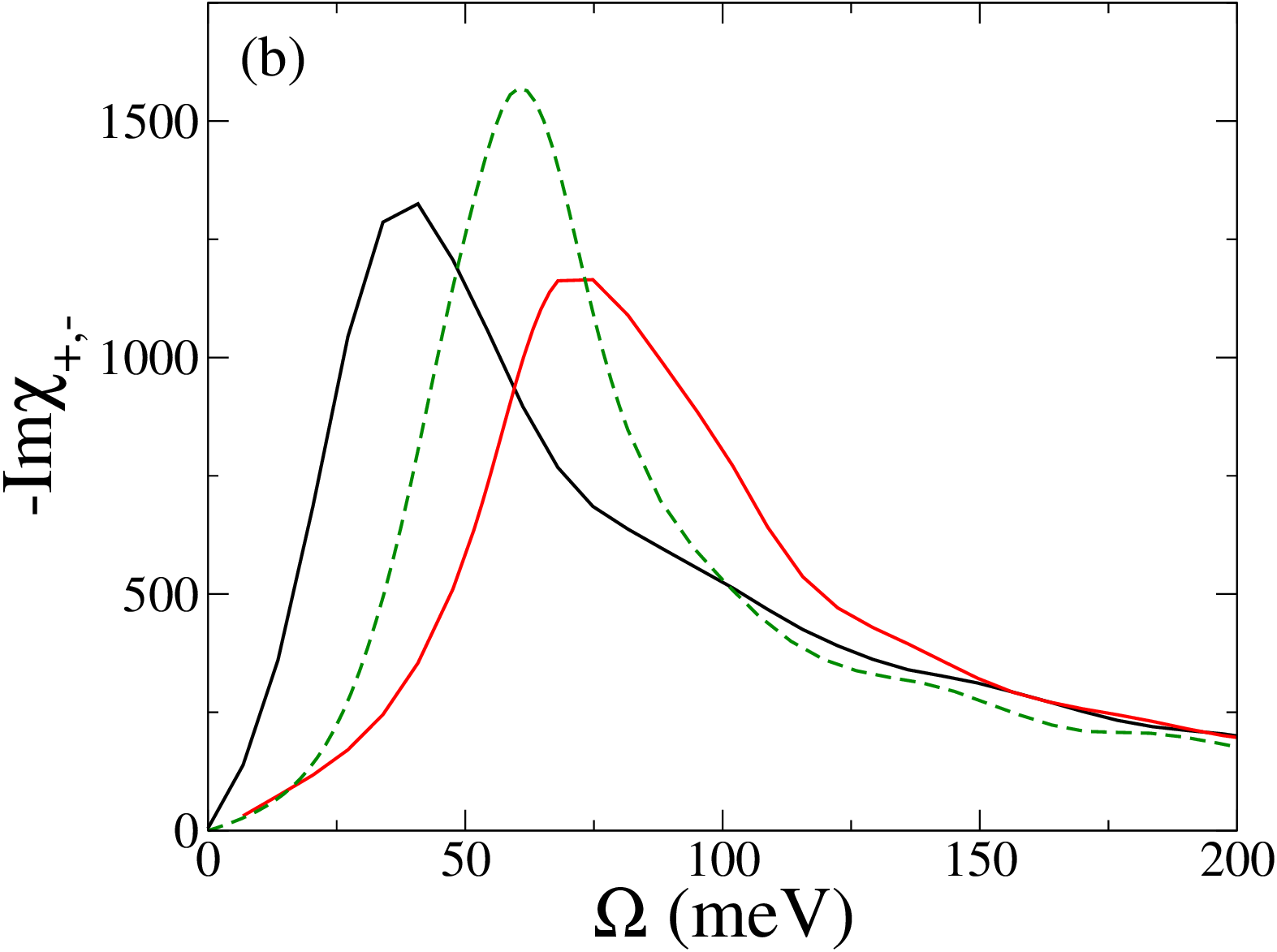}
\includegraphics[width=0.45\textwidth,clip]{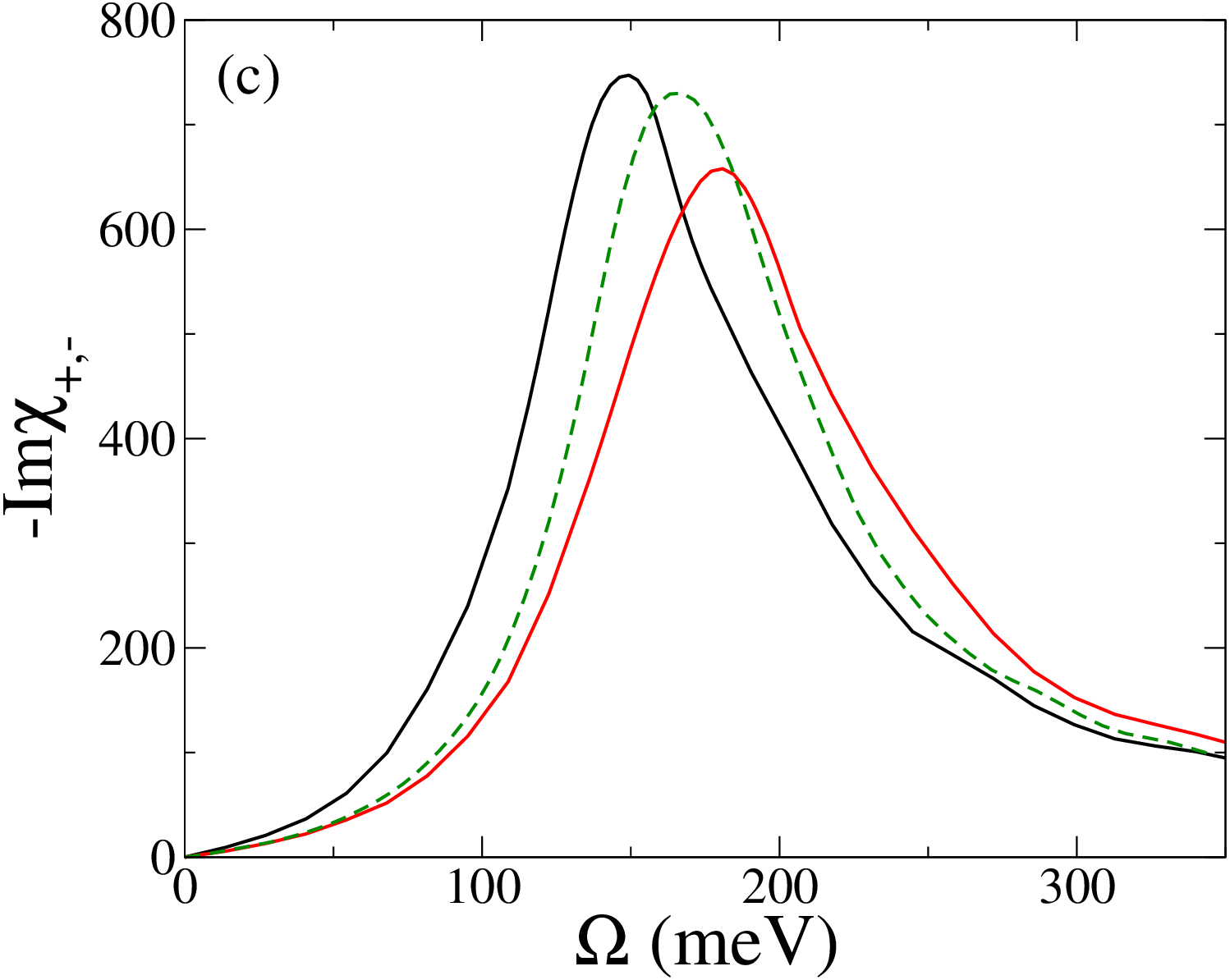}
\caption{(color online) The spectral density functions $A(\vec{Q}_\parallel,\Omega;l_\perp)$
evaluated in the Fe monolayer for three values of the wave vector in the
direction perpendicular to the magnetization. We have (a)
$|\vec{Q}_\parallel|=0.4\mathrm{\AA}^{-1}$, (b)
$|\vec{Q}_\parallel|=1.0\mathrm{\AA}^{-1}$  and
(c) $|\vec{Q}_\parallel|=1.4\mathrm{\AA}^{-1}$.
The green curve (dashed) is calculated with spin orbit coupling set to zero; the
spectral density here is independent of the sign of $\vec{Q}_\parallel$.  The
red and black curves are calculated with spin orbit coupling turned on. Now we
see asymmetries for propagation across the magnetization, with the red curve $\vec{Q}_\parallel$
directed from left to right and the black curve from right to left.}
\label{fig2}
\end{figure}

In Fig.~\ref{fig2}, for the Fe monolayer on W(110), we show the spectral density
function calculated for three values of $|\vec{Q}_\parallel|$, for propagation
across the magnetization.  Thus, the wave vector is directed along the short
axis in the surface. This is the direction probed in SPEELS studies of the Fe
monolayer on this surface.\cite{22}  In each figure, we show three curves. The
green dashed curve is calculated with spin orbit coupling set to zero. We show
only a single curve for this case, because the spectral density is identical for
the two directions of propagation across the magnetization, $+\vec{Q}_\parallel$
and  $-\vec{Q}_\parallel$. When spin orbit coupling is switched on, for the two
directions just mentioned the response function is very different, as we see
from the red and black curve in the various panels. These spin wave frequencies,
deduced from the peak in the response functions as discussed in section I,
differ for the two directions of propagation, and also note that the peak
intensities and linewidths differ as well. It is the absence of both time
reversal symmetry and reflection symmetry which renders $+\vec{Q}_\parallel$ and
$-\vec{Q}_\parallel$ inequivalent for this direction of propagation. The system
senses this breakdown of symmetry through the spin orbit interaction. If one
considers propagation parallel to the magnetization, the asymmetries displayed
in Fig.~\ref{fig2} are absent. The reason is that for this direction of
propagation, reflection in the plane that is perpendicular to both the
magnetization and the surface is a good symmetry operation of the system, but
takes  $+\vec{Q}_\parallel$ into  $-\vec{Q}_\parallel$ thus rendering the two
directions equivalent. Recall, of course, that the magnetization is a pseudo
vector in regard to reflections. Notice how very broad the curves are for large
wave vectors; the lifetime of the spin waves is very short indeed. 

\begin{figure}
\includegraphics[width=0.45\textwidth,clip]{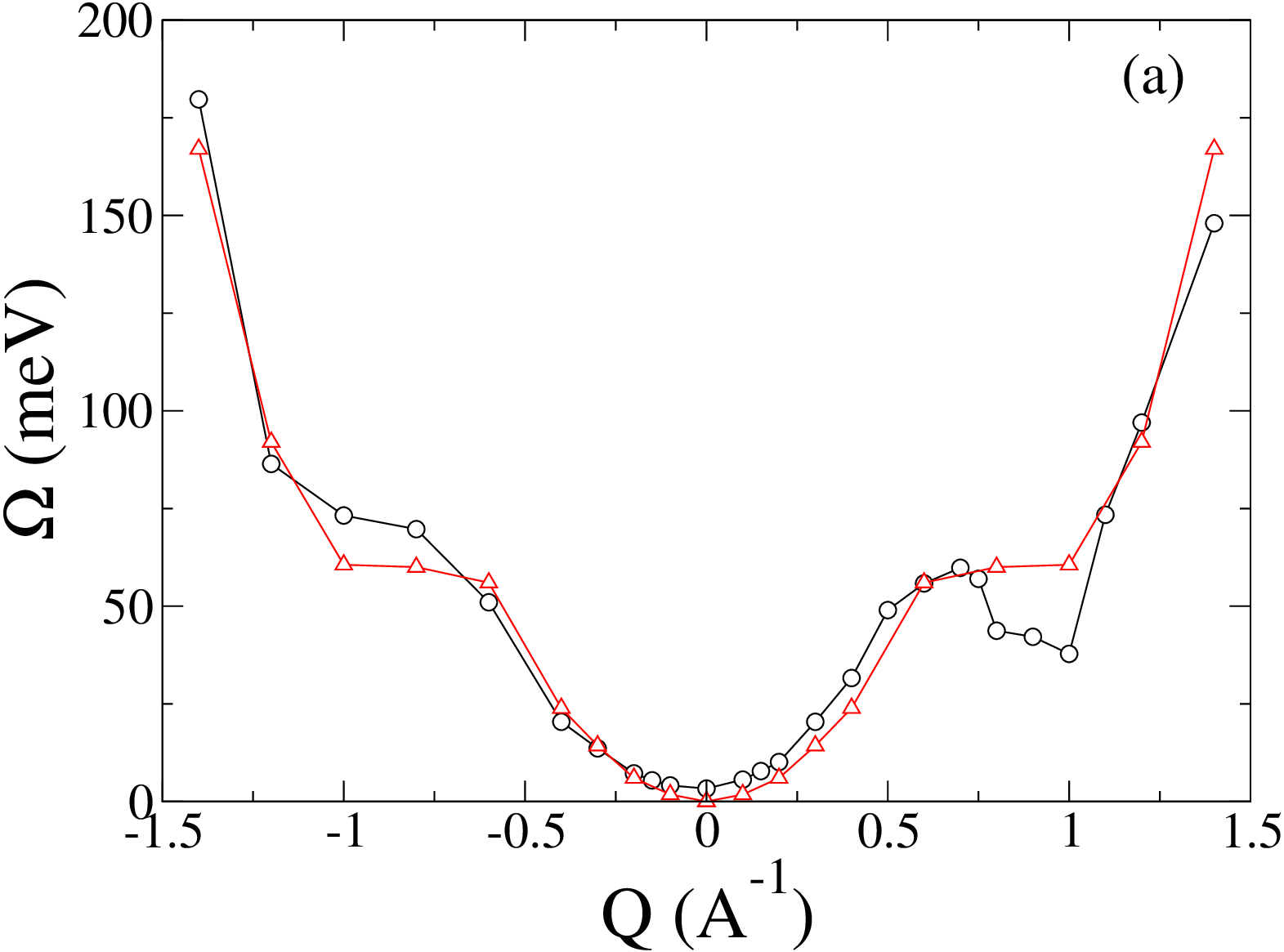}
\includegraphics[width=0.45\textwidth,clip]{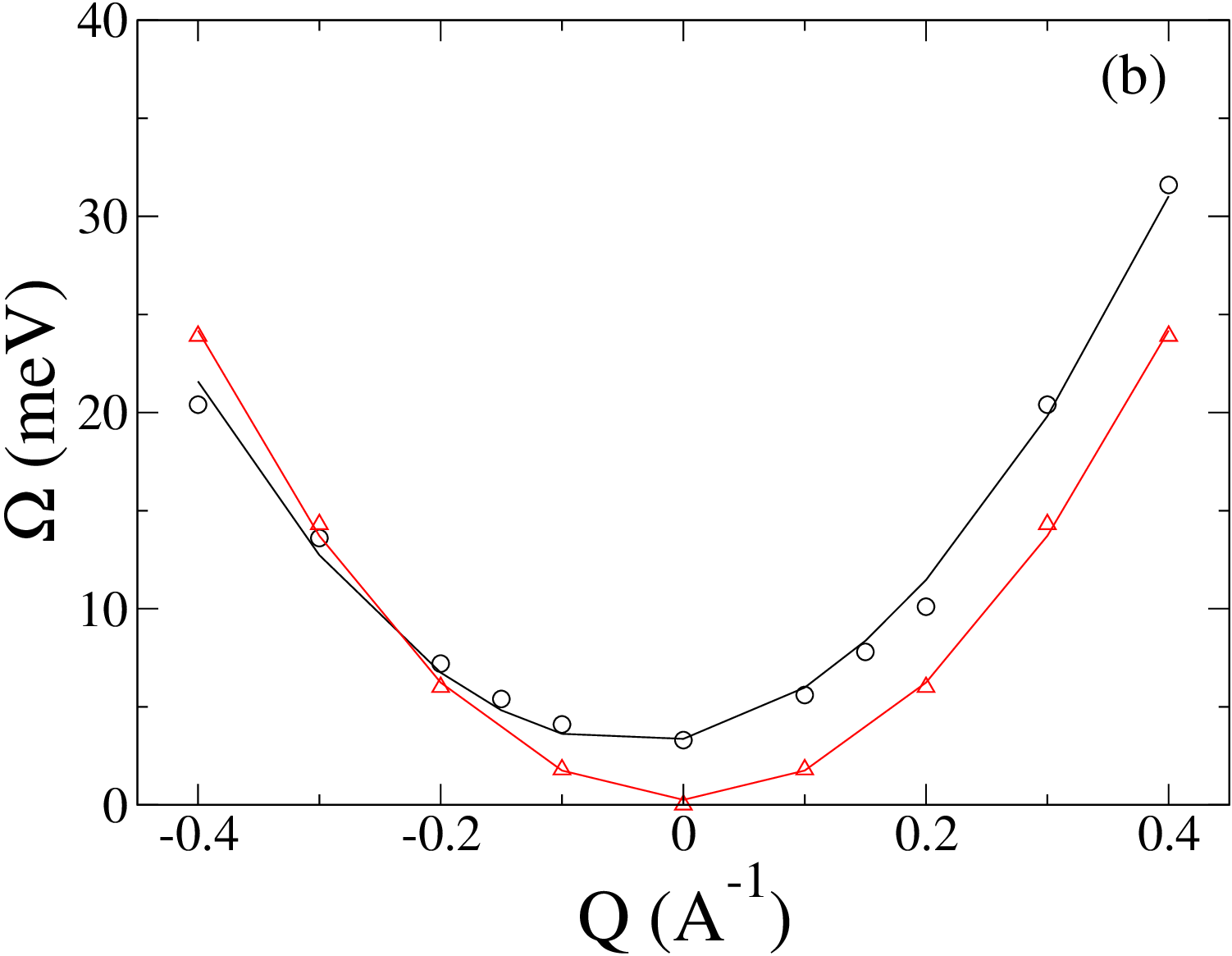}
\caption{(color online) Spin wave dispersion relations constructed from peaks in the spectral
density, for the Fe monolayer on W(110). The wave vector is in the direction
perpendicular to the magnetization. The red curve is constructed in the absence
of spin orbit coupling, it is included in the black curve.}
\label{fig3}
\end{figure}

As discussed in section I, we may construct a spin wave dispersion curve by
plotting the maxima in spectral density plots such as those illustrated in
Fig.~\ref{fig2}
as a function of wave vector. We show dispersion relations constructed in this
manner in Fig.~\ref{fig3}, with spin orbit coupling both present and absent. 
In Fig.~\ref{fig3}a,
and for propagation perpendicular to the magnetization we show the dispersion
curve so obtained for wave vectors throughout the surface Brillouin zone, and in
Fig.~\ref{fig3}b we show its behavior for small wave vectors.

Let is first consider Fig, 3(a). Here the dispersion curve extends throughout
the two dimensional Brillouin zone. At the zone boundary, quite clearly the
slope of the dispersion curve does not vanish. In this direction of propagation,
the nature of the point at the zone boundary does not require the slope to
vanish. What is most striking, clearly, is the anomaly in the vicinity of
1~$\mathrm{\AA}^{-1}$.
This feature is evident in the calculation with spin orbit coupling absent, and
for positive values of the wave vector the feature becomes much more dramatic
when spin orbit coupling is switched on. Anomalies rather similar to those in
the black curve in Fig. 3(a) appear in the green dispersion curve found in 
Fig.~3 of ref.~\onlinecite{12}, though these authors did not continue their calculation much
beyond the 1~$\mathrm{\AA}^{-1}$ regime. Our spin waves are very much softer than theirs in
this spectral region, notice. In Fig.~3 of ref.~\onlinecite{12}, one finds two dispersion
curves, one a mirror image of the second. Thus, these authors display two spin
wave frequencies for each wave vector. This surely is not correct. For a
structure with one atom per unit cell, there is one and only one spin wave mode
for each wave vector, though as discussed above for the structure explored here
symmetry allow the left/right asymmetry in the dispersion curve illustrated in
our Fig.~\ref{fig3}.

In Fig.~\ref{fig3}b, again with spin orbit coupling switched on and off, we show an
expanded view of the dispersion curve for small wave vectors. With spin orbit
interaction switched off, at zero wave vector we see a zero frequency spin wave
mode, as required by the Goldstone theorem when the underlying Hamiltonian is
form invariant under spin rotation. The curve is also symmetrical, and is
accurately fitted by the form $\Omega(\vec{Q}_\parallel)=149{Q}^2_\parallel$~(meV), 
with the wave vector in $\mathrm{\AA}^{-1}$, whereas with spin
orbit coupling turned on the dispersion relation is fitted by
$\Omega(\vec{Q}_\parallel)=3.4-11.8Q_\parallel + 143{Q}^2_\parallel$~(meV). Spin orbit
coupling introduces an anisotropy gap at $Q_\parallel=0$, and most striking is the term linear
in wave vector. This has its origin in the Dzyaloshinskii Moriya interaction
whose presence, as argued by the authors of ref.~\onlinecite{9}, has its origin in the
absence of both time reversal and inversion symmetry, for the adsorbed layer. 

At long wavelengths, one may describe spin waves by classical long wavelength
phenomenology. The linear term in the dispersion curve has its origin in a term
in the energy density of the spin system of the form
\begin{equation}
V_{DM}=-\Gamma\int dx dz S_y(x,z)\frac{\partial S_x(x,z)}{\partial x}
\end{equation}
Here $S_\alpha(x,z)$ is a spin density, the $xz$ plane is parallel to the surface,
and the magnetization is parallel to the $z$ direction. 

One interesting feature of the spin wave mode whose dispersion relation is
illustrated in Fig.~\ref{fig3}b is that at $Q_\parallel=0$, the mode has a 
finite group velocity. The fit to the dispersion curve gives this group velocity 
to be $\frac{\partial\Omega(\vec{Q}_\parallel)}{\partial Q_\parallel}\approx
2\times 10^5$cm/s , which is in the range of acoustic phonon group velocities. 

We turn now to our calculations of spin waves and the response functions for the
Fe bilayer on W(110). Let us first note that experimentally the orientation of
the magnetization in the bilayer appears to be dependent on the surface upon
which the bilayer is grown. For instance, when the bilayer is on the stepped
W(110) surface, it is magnetized perpendicular to the surface,\cite{25} a result in
agreement with ab initio calculations of the anisotropy realized in the
epitaxial bilayer.\cite{38} However, in the SPEELS studies of spin excitations in
the bilayer\cite{22,39} the magnetization is in plane. In our calculations, we
find for model B the magnetization is perpendicular to the surface, whereas in
model C it lies in plane, along the long axis very much as in the SPEELS
experiments. The anisotropy in the bilayer is not particularly large, on the
order of 0.5 meV/Fe atom, and one sees from these results that it is a property
quite sensitive to the details of the electronic structure. The fact that model
B and model C give the two different stable orientation of the magnetization
allows us to explore spin excitations for the two different orientations of the
magnetization.

\begin{figure}
\includegraphics[width=0.8\textwidth]{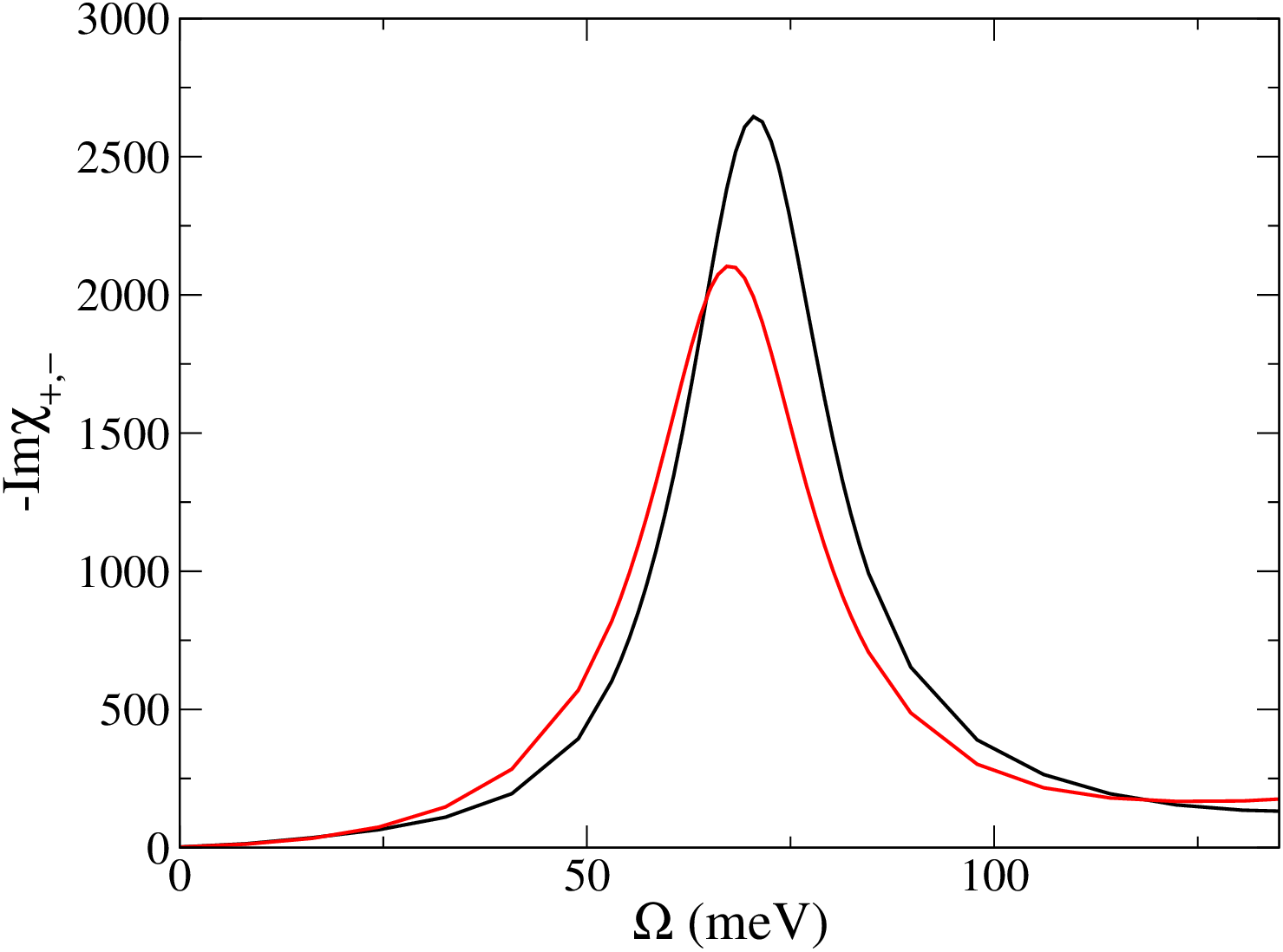}
\caption{(color online) The spectral density in the innermost layer, in the acoustic spin wave
regime, for wave vectors of $Q_\parallel = +0.5 \mathrm{\AA}^{-1}$ (black curve) and 
$Q_\parallel = -0.5 \mathrm{\AA}^{-1}$ (red curve). Model C has been used for the
calculation. In the ground state, the magnetization lies in plane along the long
axis.}
\label{fig4}
\end{figure}

We first turn our attention to the case where the magnetization lies in plane.
The bilayer has two spin wave modes, an acoustic mode for which the
magnetization in the two planes precesses in phase, and an optical mode for
which they precess 180 degrees out of phase. In Fig.~\ref{fig4}, we show calculations of
the dynamic susceptibility in the frequency range of the acoustic mode for two
values of the wave vector, $Q_\parallel=+0.5\mathrm{\AA}^-1$ and 
$Q_\parallel=-0.5\mathrm{\AA}^-1$. A spin orbit induced left right asymmetry is
clearly evident both in the peak frequency and the height of the feature. Very
recently, beautiful measurements of spin orbit asymmetries in the Fe bilayer have
appeared,\cite{39} and the results of our Fig.~\ref{fig4} are to be compared
with Fig.~3 of ref.~\onlinecite{39}. Theory and experiment are very similar,
both in regard to the intensity asymmetry and also the spin orbit induced
frequency shift, though our calculated spin wave frequencies are a little
stiffer than those found experimentally.

\begin{figure}
\includegraphics[width=0.8\textwidth]{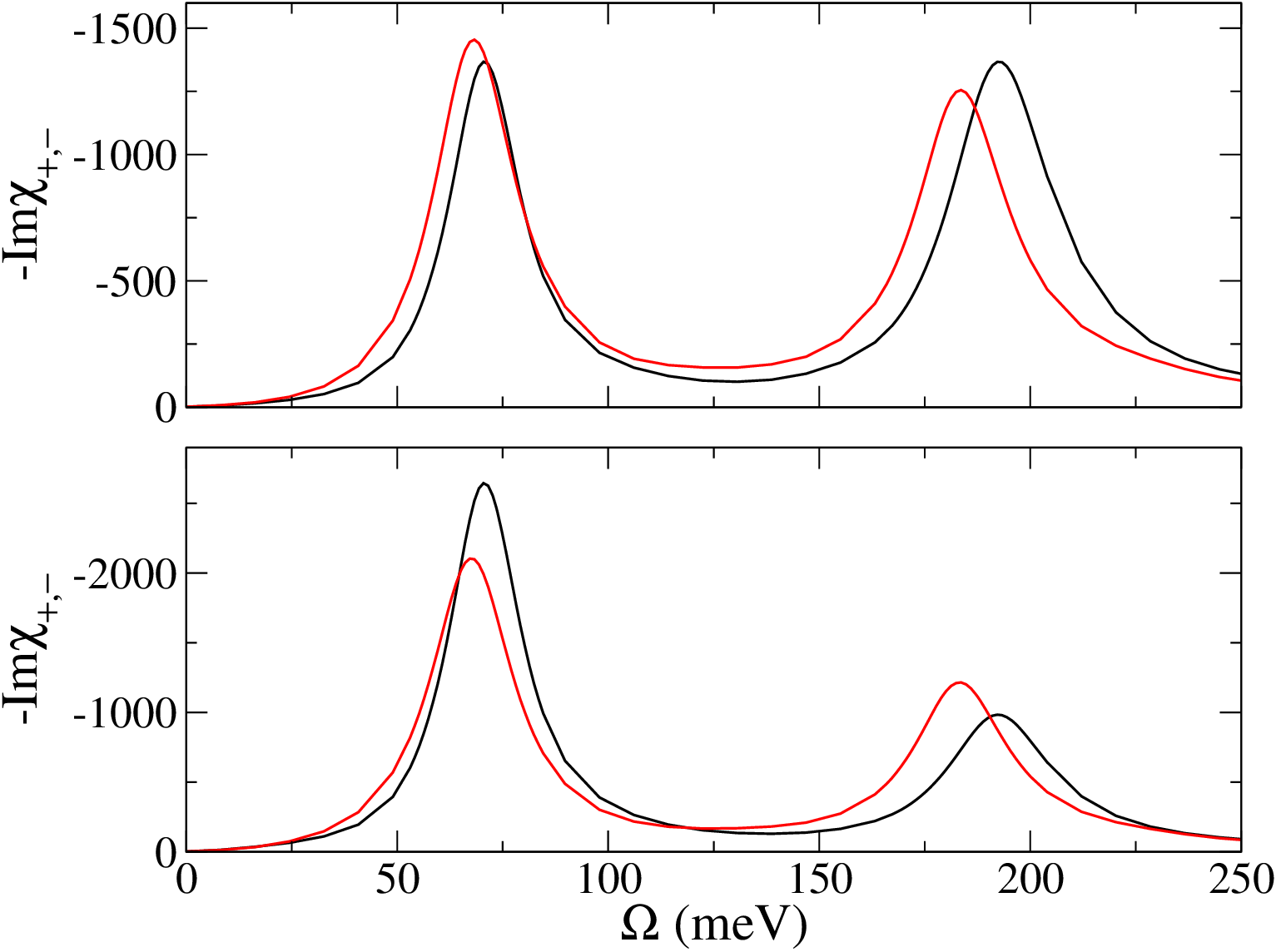}
\caption{(color online) For the wave vector $Q_\parallel = 0.5 \mathrm{\AA}^{-1}$ 
we show the spectral
densities in the innermost Fe layer (upper panel) and in the outermost layer
(lower panel) for the Fe bilayer on W(110). The figure includes the optical spin
wave feature. As in Fig.~\ref{fig4}, the black curve is calculated for
$Q_\parallel$ positive, and the red curves are for $Q_\parallel$ negative. 
The calculations employ model C.}
\label{fig5}
\end{figure}

As remarked above, in Fig.~\ref{fig4} we show only the acoustical spin wave mode
frequency regime. In Fig.~\ref{fig5}, for the spectral densities in the innermost layer
(upper panel) and the outermost layer (lower panel) we show the spectral
densities for the entire spin wave regime, including the region where the
optical spin wave is found. It is clear that the spin orbit induced frequency
shifts are largest for the optical mode which, unfortunately is not observed in
the experiments.\cite{39} 

\begin{figure}
\includegraphics[width=0.6\textwidth,clip]{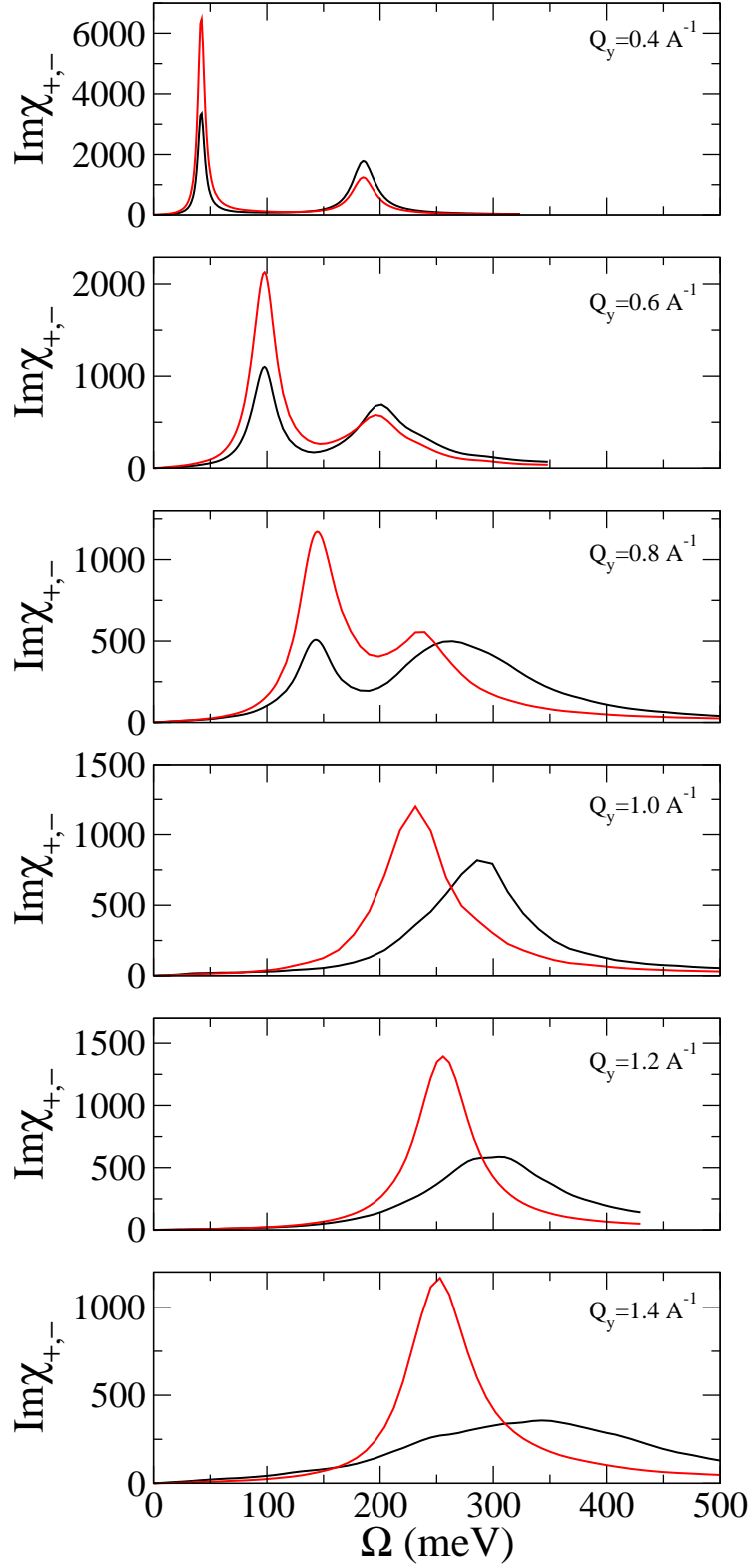}
\caption{(color online) For the Fe bilayer and for several values of the wave
vector (all positive), we show the spectral density functions for the innermost
layer adjacent to the substrate (black curve) and those for the outer layer of
the film. The calculations employ model C.} 
\label{fig6}
\end{figure}

In Fig.~\ref{fig6} for a sequence of wave vectors, all chosen positive, we show a
sequence of spectra calculated for the entire frequency range so both the
acoustic and optical spin wave feature are displayed. The black curves show the
spectral density of the innermost Fe layer, and the red curves are for the outer
layer. The optical spin wave mode, not evident in the data, shows clearly in
these figures. Notice that for wave vectors greater than 1$\mathrm{\AA}^{-1}$ the acoustical
mode is localized in the outer layer and the optical mode is localized on the
inner layer. The optical mode is very much broader than the acoustical mode at
large wave vectors, by virtue of the strong coupling to the electron hole pairs
in the W 5d bands. 

An interesting issue is the absence of the optical mode from the SPEELS spectra
reported in refs.~\onlinecite{22} and~\onlinecite{39}. We note that these spectra are taken with
only two beam energies, 4~eV and 6.75~eV. At such very low energies, the beam
electron will sample both Fe layers, so the SPEELS signal will be a coherent
superposition of electron waves backscattered from each layer; the excitation
process involves coherent excitation of both layers by the incident electron.
As a consequence of the 180 degree phase difference in spin motions associated
with the two modes it is quite possible, indeed even probable, that for energies
where the acoustical mode is strong the intensity of the optical mode is weak,
by virtue of quantum interference effects in the excitation scattering
amplitude. In earlier studies of surface phonons, it is well documented that on
surfaces where two surface phonons of different polarization exist for the same
wave vector, one can be silent and one active in electron loss
spectroscopy.\cite{40} 
It would require a full multiple scattering analysis of the spin wave
excitation process to explore this theoretically. While earlier\cite{41}
calculations that address SPEELS excitation of spin waves described by the
Heisenberg model could be adapted for this purpose, in principle, a problem is
that at such low beam energies it is necessary to take due account of image
potential effects to obtain meaningful results.\cite{42} This is very difficult to
do without  considerable information on the electron reflectivity of the 
surface.\cite{42} It would be of great interest to see experimental SPEELS studies of the Fe
bilayer with a wider range of beam energies to search for the optical mode, if
this were possible.  

\begin{figure}
\includegraphics[width=0.6\textwidth]{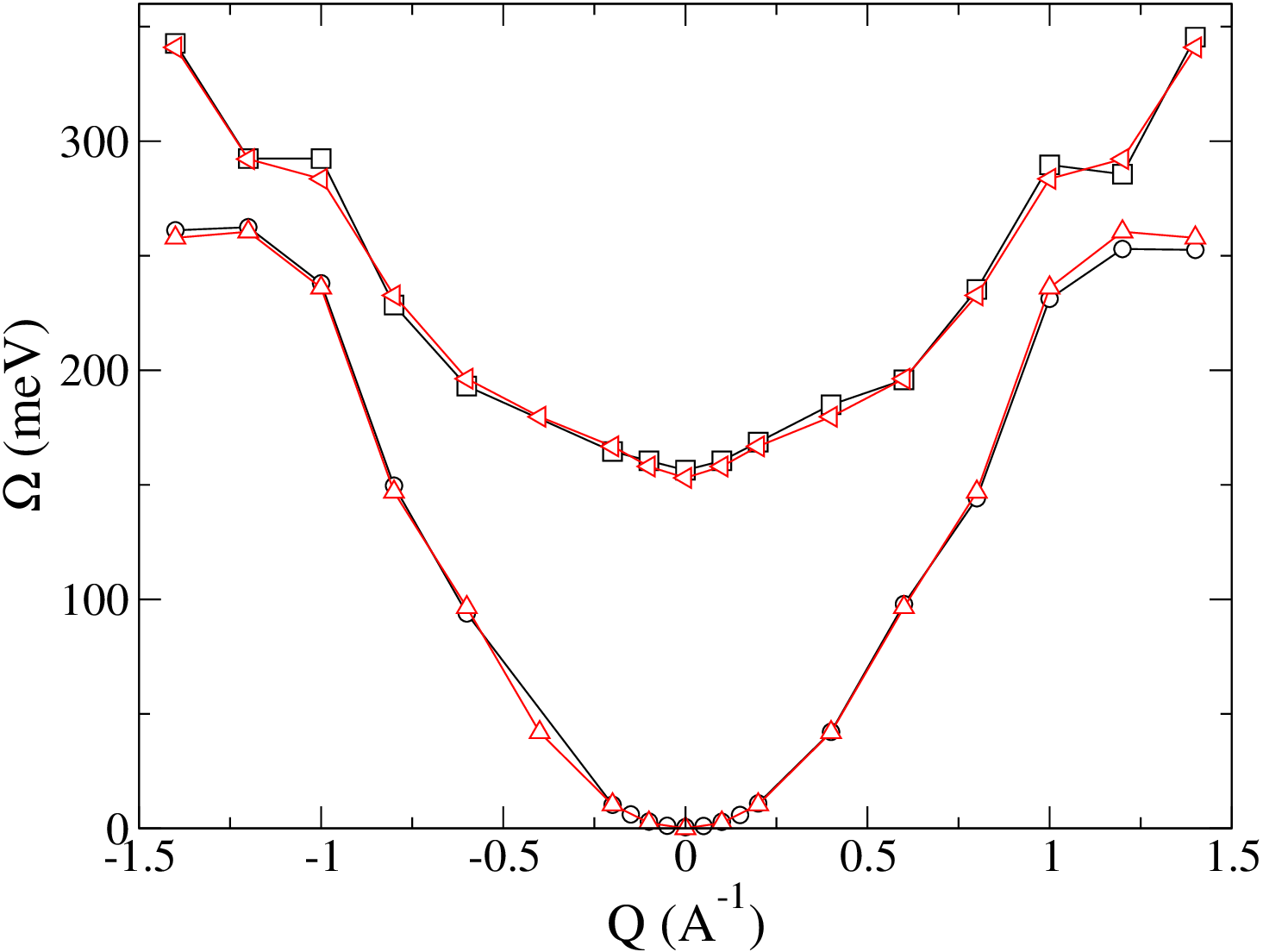}
\caption{(color online) For the Fe bilayer with magnetization in plane, we show
the spin wave dispersion curves calculated with spin orbit coupling (black
points) and without spin orbit coupling (red curves). Model C has been employed
for these calculations.}
\label{fig7} 
\end{figure}

In Fig.~\ref{fig7}, we show dispersion curves for the optical and acoustic spin wave
branches for the bilayer. The magnetization lies in plane, and one can see that
on the scale of this figure, the spin orbit effects on the dispersion curve are
rather modest compared to those in the monolayer. For small wave vectors, with
spin orbit coupling present, the dispersion curve of the acoustic spin wave
branch is fitted by the form $\Omega(Q_\parallel)=0.49-0.85 Q_\parallel+243
Q^2_\parallel$(meV) so at long wavelengths the 
influence of the Dzyaloshinskii Moriya interaction is more than one order of 
magnitude smaller than it is in the monolayer. 

If the magnetization is perpendicular to the surface, then symmetry
considerations show that there are no left/right asymmetries in the spin wave
propagation characteristics. One may see this as follows. Consider a wave vector
$\vec{Q}_\parallel$ in the plane of the surface, which also is perpendicular to
the magnetization, and thus perpendicular to the long axis. The reflection $R$
in the plane perpendicular to the surface and which contains the magnetization
simultaneously changes the sign of wave vector and the magnetization. If this is
followed by the time reversal operation $T$, then $\vec{Q}_\parallel$ remains
reversed in sign but the magnetization changes back to its original orientation.
Thus the product $RT$ leaves the system invariant but transforms
$\vec{Q}_\parallel$ into  $-\vec{Q}_\parallel$. The two propagation directions
are then equivalent.

\begin{figure}
\includegraphics[width=0.6\textwidth]{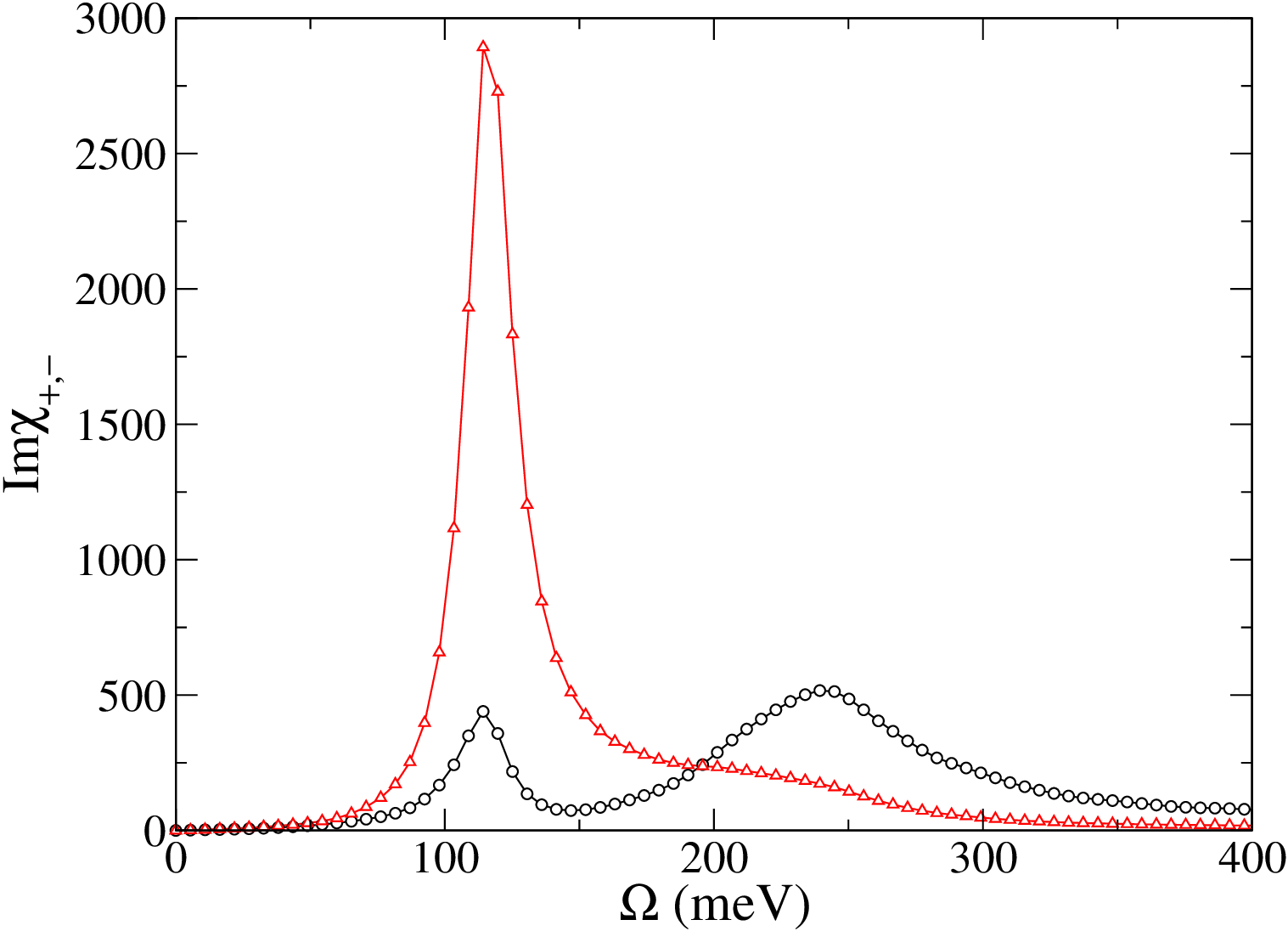}
\caption{(color online) For the bilayer and the case where the magnetization is
perpendicular to the surface (model B), and for $Q_\parallel=0.6\mathrm{\AA}^{-1}$, 
we show spectral density function calculated for positive
values of $Q_\parallel$  (continuous lines) and negative values of $Q_\parallel$
(symbols). The black curve is the spectral density for the inner layer, and the
red curve is the outermost Fe layer.} 
\label{fig8}
\end{figure}

We illustrate this in Fig.~\ref{fig8} where, for $Q_\parallel=0.6\mathrm{\AA}^{-1}$, 
where it is shown that the
spectral densities calculated for the two directions of propagation are
identical, with spin orbit coupling switched on. Model B, in which the
magnetization is perpendicular to the surface, has been used in these
calculations. The spectral densities calculated for the two signs of
$Q_\parallel$ cannot
be distinguished to within the numerical precision we use. 

\section{Concluding Remarks}

We have developed the formalism which allows one to include the influence of
spin orbit coupling on the spin excitations of ultrathin ferromagnets on semi
infinite metallic substrates. Our approach allows us to calculate the full
dynamic susceptibility of the system, so as illustrated by the calculations
presented in section III we can examine the influence of spin orbit coupling on
the linewidth (or lifetime) of spin excitations, along with their oscillator
strength. As in previous work, we can then construct effective dispersion curves
by following peaks in the spectral density as a function of wave vector, without
resort to calculations of large numbers of very small distant neighbor exchange
interactions.  The results presented in Fig.~\ref{fig4} are very similar to the
experimental data reported in ref.~\onlinecite{39}, as discussed above, though we see that
in the bilayer the influence of the Dzyaloshinskii Moriya interaction is
considerably more modest than in the monolayer. 

We will be exploring other issues in the near future. One interest in our minds
is the influence of spin orbit coupling on the spin pumping contributions to the
ferromagnetic resonance linewidth, as observed in ferromagnetic resonance (FMR)
studies of ultrathin films.\cite{43} It has been shown earlier\cite{44} that the
methodology employed in the present paper (without spin orbit coupling included)
can be applied to the description of the spin pumping contribution to the FMR
linewidth, and in fact an excellent quantitative account of the data on the
Fe/Au(100) system was obtained. It is possible that for films grown on 4d and 5d
substrates that spin orbit coupling can influence the spin pumping relaxation
rate substantially. This will require calculations directed toward much thicker
films than explored here. The formalism we have developed and described in the
present paper will allow such studies in the future. 

\acknowledgments

This research was supported by the U. S. Department of Energy, through grant No.
DE-FG03-84ER-45083. S. L. wishes to thank the Alexander von Humboldt Foundation
for a Feodor Lynen Fellowship. A.T.C. and R.B.M. acknowledge support from CNPq
and FAPERJ and A.B.K. was supported also by the CNPq, Brazil. 

\appendix*

\section{}

In this Appendix we provide explicit expressions for the various quantities
which enter the equations displayed in Section II. While these expressions are
unfortunately lengthy, it will be useful for them to be given in full.

\begin{eqnarray}
	A^{(1)}_{\mu\nu,\mu'\nu'}(ll';mm') = 
	\delta_{l'm}\delta_{\nu\mu'}\langle c^\dagger_{l\mu\up}c_{m'\nu'\up}\rangle - 
	\delta_{lm'}\delta_{\mu\nu'}\langle c^\dagger_{m\mu'\dn}c_{l'\nu\dn}\rangle \nonumber\\
	A^{(2)}_{\mu\nu,\mu'\nu'}(ll';mm') = 
	-\delta_{lm'}\delta_{\mu\nu'}\langle c^\dagger_{m\mu\dn}c_{l'\nu\up}\rangle \nonumber\\
	A^{(3)}_{\mu\nu,\mu'\nu'}(ll';mm') = 
	\delta_{l'm}\delta_{\nu\mu'}\langle 
	c^\dagger_{l\mu\dn}c_{m'\nu'\up}\rangle  
	\nonumber\\
	A^{(4)}_{\mu\nu,\mu'\nu'}(ll';mm')  = 0
\label{As}
\end{eqnarray}

The various expectation values in the equations above and those displayed below
are calculated from the single particle Green’s functions once the self
consistent ground state parameters are determined. Then

\begin{eqnarray}
	\tilde{B}^{11}_{\mu\nu,\mu'\nu'}(ll';mm') = 
	\sum_{\eta}\left( U_{l;\mu'\eta,\mu\nu'}\langle 
	c^\dagger_{l\eta\dn}c_{l'\nu\dn}\rangle\delta_{lm}\delta_{lm'} -
	U_{l';\mu'\nu,\eta\nu'}\langle c^\dagger_{l\mu\up}c_{l'\eta\up}\rangle\delta_{l'm}\delta_{l'm'}
	\right) \nonumber\\
	\tilde{B}^{12}_{\mu\nu,\mu'\nu'}(ll';mm') = 
	\sum_{\eta}\left( U_{l';\mu'\nu,\nu'\eta}\langle 
	c^\dagger_{l\mu\up}c_{l'\eta\dn}\rangle\delta_{l'm}\delta_{l'm'} -
	U_{l;\eta\mu',\mu\nu'}\langle 
	c^\dagger_{l\eta\up}c_{l'\nu\dn}\rangle\delta_{lm}\delta_{lm'} +\right. \nonumber\\ 
	\left. + U_{l;\mu'\eta,\mu\nu'}\langle 
	c^\dagger_{l\eta\up}c_{l'\nu\dn}\rangle\delta_{lm}\delta_{lm'} 
	\right) \nonumber\\
	\tilde{B}^{13}_{\mu\nu,\mu'\nu'}(ll';mm') = 
	\sum_{\eta}\left( U_{l';\mu'\nu,\nu'\eta}\langle 
	c^\dagger_{l\mu\up}c_{l'\eta\dn}\rangle\delta_{l'm}\delta_{l'm'} -
	U_{l;\eta\mu',\mu\nu'}\langle 
	c^\dagger_{l\eta\up}c_{l'\nu\dn}\rangle\delta_{lm}\delta_{lm'} -\right. \nonumber\\ 
	\left. -U_{l';\mu'\nu,\eta\nu'}\langle 
	c^\dagger_{l\mu\up}c_{l'\eta\dn}\rangle\delta_{l'm}\delta_{l'm'} 
	\right) \nonumber\\
	\tilde{B}^{14}_{\mu\nu,\mu'\nu'}(ll';mm')=0\nonumber\\
\label{Bts_1}
\end{eqnarray}

\begin{eqnarray}
	\tilde{B}^{21}_{\mu\nu,\mu'\nu'}(ll';mm') = 
	\sum_{\eta} U_{l;\mu'\eta,\mu\nu'}\langle
	c^\dagger_{l\eta\dn}c_{l'\nu\up}\rangle\delta_{lm}\delta_{lm'}\nonumber\\
 	\tilde{B}^{22}_{\mu\nu,\mu'\nu'}(ll';mm') = \sum_{\eta}\left[ 
	\left(U_{l';\mu'\nu,\nu'\eta}-U_{l';\mu'\nu,\eta\nu'}\right)
	\langle c^\dagger_{l\mu\up}c_{l'\eta\up}\rangle\delta_{l'm}\delta_{l'm'}-
	\right. \nonumber\\
	\left. - \left(U_{l;\eta\mu',\mu\nu'}-U_{l;\mu'\eta,\mu\nu'}\right)
	\langle c^\dagger_{l\eta\up}c_{l'\nu\up}\rangle\delta_{lm}\delta_{lm'}\right]\nonumber\\
	\tilde{B}^{23}_{\mu\nu,\mu'\nu'}(ll';mm') = \sum_{\eta}\left(
	 U_{l;\mu'\nu,\nu'\eta}
        \langle c^\dagger_{l\mu\up}c_{l'\eta\up}\rangle\delta_{l'm}\delta_{l'm'}- 
	U_{l;\eta\mu',\mu\nu'}
        \langle c^\dagger_{l\eta\up}c_{l'\nu\up}\rangle\delta_{lm}\delta_{l'm'}\right)\nonumber\\
	\tilde{B}^{24}_{\mu\nu,\mu'\nu'}(ll';mm') = 
	-\sum_{\eta} U_{l';\mu'\nu,\eta\nu'}\langle
	c^\dagger_{l\mu\up}c_{l'\eta\dn}\rangle\delta_{l'm}\delta_{l'm'}\nonumber\\
\label{Bts_2}
\end{eqnarray}

\begin{eqnarray}
	\tilde{B}^{31}_{\mu\nu,\mu'\nu'}(ll';mm') = 
	-\sum_{\eta} U_{l';\mu'\nu,\eta\nu'}\langle
	c^\dagger_{l\mu\dn}c_{l'\eta\up}\rangle\delta_{l'm}\delta_{l'm'}\nonumber\\
	\tilde{B}^{32}_{\mu\nu,\mu'\nu'}(ll';mm') = \sum_{\eta}\left(
	 U_{l;\mu'\nu,\nu'\eta}
        \langle c^\dagger_{l\mu\dn}c_{l'\eta\dn}\rangle\delta_{l'm}\delta_{l'm'}- 
	U_{l;\eta\mu',\mu\nu'}
        \langle c^\dagger_{l\eta\dn}c_{l'\nu\dn}\rangle\delta_{lm}\delta_{l'm'}\right)\nonumber\\
	\tilde{B}^{33}_{\mu\nu,\mu'\nu'}(ll';mm') = \sum_{\eta}\left[ 
	\left(U_{l';\mu'\nu,\nu'\eta}-U_{l';\mu'\nu,\eta\nu'}\right)
	\langle c^\dagger_{l\mu\dn}c_{l'\eta\dn}\rangle\delta_{l'm}\delta_{l'm'}-
	\right. \nonumber\\
	\left. -\left(U_{l;\eta\mu',\mu\nu'}-U_{l;\mu'\eta,\mu\nu'}\right)
	\langle c^\dagger_{l\eta\dn}c_{l'\nu\dn}\rangle\delta_{lm}\delta_{lm'}\right]\nonumber\\
	\tilde{B}^{34}_{\mu\nu,\mu'\nu'}(ll';mm') = 
	\sum_{\eta} U_{l;\mu'\eta,\mu\nu'}\langle
	c^\dagger_{l\eta\up}c_{l'\nu\dn}\rangle\delta_{lm}\delta_{lm'}\nonumber\\
\label{Bts_3}
\end{eqnarray}

\begin{eqnarray}
	\tilde{B}^{41}_{\mu\nu,\mu'\nu'}(ll';mm') = 0\nonumber\\
	\tilde{B}^{42}_{\mu\nu,\mu'\nu'}(ll';mm') = 
	\sum_{\eta}\left( U_{l';\mu'\nu,\nu'\eta}\langle 
	c^\dagger_{l\mu\dn}c_{l'\eta\up}\rangle\delta_{l'm}\delta_{l'm'} -
	U_{l;\eta\mu',\mu\nu'}\langle 
	c^\dagger_{l\eta\dn}c_{l'\nu\up}\rangle\delta_{lm}\delta_{lm'} -\right. \nonumber\\
	\left. -U_{l';\mu'\nu,\eta\nu'}\langle 
	c^\dagger_{l\mu\dn}c_{l'\eta\up}\rangle\delta_{l'm}\delta_{l'm'} 
	\right) \nonumber\\
	\tilde{B}^{43}_{\mu\nu,\mu'\nu'}(ll';mm') = 
	\sum_{\eta}\left( U_{l';\mu'\nu,\nu'\eta}\langle 
	c^\dagger_{l\mu\dn}c_{l'\eta\up}\rangle\delta_{l'm}\delta_{l'm'} -
	U_{l;\eta\mu',\mu\nu'}\langle 
	c^\dagger_{l\eta\dn}c_{l'\nu\up}\rangle\delta_{lm}\delta_{lm'} +\right. \nonumber\\ 
	\left. +U_{l;\mu'\eta,\mu\nu'}\langle 
	c^\dagger_{l\eta\dn}c_{l'\nu\up}\rangle\delta_{lm}\delta_{lm'} 
	\right) \nonumber\\
	\tilde{B}^{44}_{\mu\nu,\mu'\nu'}(ll';mm') = 
	\sum_{\eta}\left( U_{l;\mu'\eta,\mu\nu'}\langle
	c^\dagger_{l\eta\up}c_{l'\nu\up}\rangle\delta_{lm}\delta_{l'm} -
	U_{l';\mu'\nu,\eta\nu'}\langle c^\dagger_{l\mu\dn}c_{l'\eta\dn}
	\rangle\delta_{l'm}\delta_{l'm'}\right)\nonumber\\
\label{Bts_4}
\end{eqnarray}

\begin{eqnarray}
	B^{11}_{\mu\nu,\mu'\nu'}(ll';mm') = 
	\tilde{T}^{\nu\nu'\dn}_{l'm'}\delta_{lm}\delta_{\mu\mu'}-
	(\tilde{T}^{\mu\mu'\up}_{lm})^*\delta_{l'm'}\delta_{\nu\nu'}\nonumber\\
	B^{12}_{\mu\nu,\mu'\nu'}(ll';mm') = 
	\alpha_{l';\nu'\nu}^*\delta_{lm}\delta_{l'm'}\delta_{\mu\mu'}\nonumber\\
	B^{13}_{\mu\nu,\mu'\nu'}(ll';mm') = 
	-\alpha_{l;\mu\mu'}^*\delta_{lm}\delta_{l'm'}\delta_{\nu\nu'}\nonumber\\
	B^{14}_{\mu\nu,\mu'\nu'}(ll';mm') = 0
\label{Bs_1}
\end{eqnarray}

\begin{eqnarray}
	B^{21}_{\mu\nu,\mu'\nu'}(ll';mm') = 
	\alpha_{l';\nu\nu'}\delta_{lm}\delta_{l'm'}\delta_{\mu\mu'}\nonumber\\
	B^{22}_{\mu\nu,\mu'\nu'}(ll';mm') = 
	\tilde{T}^{\nu\nu'\up}_{l'm'}\delta_{lm}\delta_{\mu\mu'}-
	(\tilde{T}^{\mu\mu'\up}_{lm})^*\delta_{l'm'}\delta_{\nu\nu'}\nonumber\\
	B^{23}_{\mu\nu,\mu'\nu'}(ll';mm') = 0\nonumber\\
	B^{24}_{\mu\nu,\mu'\nu'}(ll';mm') = 
	-\alpha_{l;\mu\mu'}^*\delta_{lm}\delta_{l'm'}\delta_{\nu\nu'}
\label{Bs_2}
\end{eqnarray}

\begin{eqnarray}
	B^{31}_{\mu\nu,\mu'\nu'}(ll';mm') = 
	-\alpha_{l;\mu'\mu}\delta_{lm}\delta_{l'm'}\delta_{\nu\nu'}\nonumber\\
	B^{32}_{\mu\nu,\mu'\nu'}(ll';mm') = 0\nonumber\\
	B^{33}_{\mu\nu,\mu'\nu'}(ll';mm') = 
	\tilde{T}^{\nu\nu'\dn}_{l'm'}\delta_{lm}\delta_{\mu\mu'}-
	(\tilde{T}^{\mu\mu'\dn}_{lm})^*\delta_{l'm'}\delta_{\nu\nu'}\nonumber\\
	B^{34}_{\mu\nu,\mu'\nu'}(ll';mm') = 
	\alpha_{l';\nu'\nu}^*\delta_{lm}\delta_{l'm'}\delta_{\mu\mu'}
\label{Bs_3}
\end{eqnarray}

\begin{eqnarray}
	B^{41}_{\mu\nu,\mu'\nu'}(ll';mm') = 0\nonumber\\
	B^{42}_{\mu\nu,\mu'\nu'}(ll';mm') = 
	-\alpha_{l;\mu'\mu}\delta_{lm}\delta_{l'm'}\delta_{\nu\nu'}\nonumber\\
	B^{43}_{\mu\nu,\mu'\nu'}(ll';mm') = 
	\alpha_{l';\nu\nu'}\delta_{lm}\delta_{l'm'}\delta_{\mu\mu'}\nonumber\\
	B^{11}_{\mu\nu,\mu'\nu'}(ll';mm') = 
	\tilde{T}^{\nu\nu'\up}_{l'm'}\delta_{lm}\delta_{\mu\mu'}-
	(\tilde{T}^{\mu\mu'\dn}_{lm})^*\delta_{l'm'}\delta_{\nu\nu'}
\label{Bs_4}
\end{eqnarray}


\begin{thebibliography}{99}
\bibitem{6} For an explicit comparison between the predictions of the
Heisenberg model and that of an itinerant electron description of the spin
excitations, see Fig.~4 and Fig.~5 of the theoretical paper cited in
ref.~\onlinecite{4}.
\end{thebibliography}
\end{document}